\documentclass[prd,aps,amsmath,amssymb,letter]{revtex4}
\usepackage{graphicx}
\usepackage{epsfig,latexsym}
\begin{document}
\title{Baryon Inhomogeneity Generation in the Quark-Gluon Plasma Phase}
\author{Biswanath Layek $^1$} 
\email {layek@veccal.ernet.in} 
\author{Ananta P. Mishra $^2$}
\email {apmishra@iopb.res.in}
\author{Ajit M. Srivastava $^2$}
\email{ajit@iopb.res.in}
\author{Vivek K. Tiwari $^3$}
\email{vivek_krt@hotmail.com}
\affiliation{$^1$Variable Energy Cyclotron Center, Kolkata 700064, India \\
$^2$Institute of Physics, Sachivalaya Marg, Bhubaneswar 751005, India \\
$^3$Physics Department, Allahabad University, Allahabad 211002, India}

\begin{abstract}

 We discuss the possibility of generation of baryon inhomogeneities
in a quark-gluon plasma phase due to moving $Z(3)$ interfaces. By
modeling the dependence of effective mass of the quarks on the
Polyakov loop order parameter, we study the reflection of quarks
from collapsing $Z(3)$ interfaces and estimate resulting baryon
inhomogeneities in the context of the early universe. We argue that 
in the context of certain low energy scale inflationary models,
it is possible that large $Z(3)$ walls arise at the end of the reheating
stage. Collapse of such walls could lead to baryon inhomogeneities which
may be separated by large distances near the QCD scale. Importantly, 
the generation of these inhomogeneities is insensitive to the order, 
or even the existence, of the quark-hadron phase transition. We also 
briefly discuss the possibility of formation of quark nuggets in this 
model, as well as baryon inhomogeneity generation in 
relativistic heavy-ion collisions. 
\end{abstract}
\maketitle

\section{Introduction}

Generation of baryon inhomogeneities in the early universe can
have important implications for nucleosynthesis, and for the possibility
of creating compact baryon rich objects \cite{wtn}. Though, current
observations do not support any strong deviation from the
standard big-bang nucleosynthesis  calculations. Calculations of
inhomogeneous big bang nucleosynthesis resulting from an
inhomogeneous distribution of baryons in the universe,
(such as those in ref. \cite{ibbn1,ibbn2}), therefore, can be used to
constrain the baryon inhomogeneities present in the early universe.

There have been numerous investigations of the nature of baryon
inhomogeneities generated during a first order quark-hadron phase transition
\cite{wtn,fuller}. In these investigations, baryon inhomogeneities
arise due to moving bubble walls at the transition, with baryons getting
concentrated in the remaining localized quark-gluon plasma (QGP) regions.

 Main problems in implementing the scenario of ref. \cite{wtn} have been
regarding the nature of the quark-hadron phase transition as well as
the relevant length scales. Lattice calculations \cite{ltc}
tell us that for realistic values of quark masses, quark-hadron
transition is at best a weak first order transition, and most likely
it is a cross-over. The scenario of ref. \cite{wtn} does not work in
this case. Even if one allows for a possibility of strong first order
transition, relevant length and time scales are such that the
resulting baryon inhomogeneities are separated by very small
distances. Typical separation between such baryonic lumps is
of the order of separation between the nucleation sites of the hadronic
bubbles, which is at most of the order of few cm at the end of the
quark-hadron transition for homogeneous nucleation\cite{fuller,impur}.
In order that these baryonic lumps survive various dissipative processes,
this separation needs to be at least of order of a meter at the transition 
stage \cite{bfluct}. There have been discussions of larger separations 
between baryon inhomogeneities invoking impurity induced inhomogeneous 
bubble nucleation \cite{impur}, presence of density fluctuations 
\cite{fluc,layek} etc. However, all these scenarios still depend 
crucially on the assumption of a first order phase transition, 
and will not work if the quark-hadron transition was a cross-over.

 In this paper we propose a different scenario where baryon inhomogeneities
are produced not due to moving quark-hadron phase boundaries, but due
to moving $Z(N)$ interfaces. $Z(N)$ interfaces arise when one uses
the expectation value of the Polyakov loop, $l(x)$, as the order parameter
for the confinement-deconfinement phase transition of an SU(N) gauge
theory \cite{plkv}.  This order parameter transforms non-trivially under
the center $Z(N)$ of the  SU(N) group  and is non-zero
above the critical temperature $T_c$. This breaks the global $Z(N)$
symmetry spontaneously above $T_c$, while the symmetry is restored
below $T_c$ in the confining phase where this order parameter vanishes.
For QCD with SU(3) color group, spontaneous breaking of the 
discrete $Z(3)$ symmetry in the QGP phase leads to the existence of 
domain walls (interfaces) across which $l(x)$ interpolates between 
different $Z(3)$ vacua.  The properties and physical consequences of
these $Z(3)$ interfaces have been discussed in the literature \cite{zn}.
Though, we mention that it has also been suggested  that these
interfaces should not be taken as physical objects in the Minkowski
space \cite{smlg}. Similarly, it has also been subject of discussion
whether it makes sense to talk about this $Z(3)$ symmetry in the presence
of quarks \cite{qurk1}. The presence of quarks can be interpreted as
leading to explicit breaking of $Z(3)$ symmetry, lifting the degeneracy
of different $Z(3)$ vacua \cite{qurk2,psrsk,psrsk2,veff}. In this approach,
with quarks, $Z(3)$ interfaces become unstable and move away from the
region with the unique true vacuum. Thus, in the context of cosmology,
if these walls were produced at some early stage (say after GUT 
scale inflation), it is likely that they will quickly disappear due 
to this pressure difference between different $Z(3)$ vacua. However,
we will argue (in section III) that in the context of certain low
energy scale inflationary models it is possible that large $Z(3)$
domain walls may arise in the QGP phase near the quark-hadron transition 
stage and may lead to observational effects.

  The basic idea of our model is that as $l(x)$ is the order parameter for
the quark-hadron transition, physical properties such as effective mass
of the quarks should be determined in terms of $l(x)$.  This also looks
natural from the expected correlation between the chiral condensate and
the Polyakov loop.  Thus, if there is
spatial variation in the value of $l(x)$ in the QGP phase then effective
mass of the quark traversing that region should also vary. For regions
where $l(x) = 0$, quarks should acquire constituent mass as appropriate
for the confining phase. As we will see below, $l(x)$ varies across
a $Z(3)$ interface, acquiring small magnitude in the center of the wall.
A quark passing through this interface, therefore, experiences a nonzero
potential barrier leading to non-zero reflection coefficient for
the quark. Due to this, as a closed domain wall collapses, quarks inside 
will stream through it.  With a non-zero reflection coefficient, net 
baryon number density inside will grow, somewhat  in the manner as in 
the conventional treatments of collapsing quark-hadron phase boundaries. 
This will lead to formation of baryonic lumps.

 Important thing to realize is that all this happens in the QGP phase
itself, with any possible quark-hadron transition being completely
irrelevant to this discussion. The only relevance of the quark-hadron
transition is that in the hadronic phase $l(x) = 0$ so all $Z(3)$
domain walls disappear. The final structure of the baryon inhomogeneities
will therefore be decided by those $Z(3)$ interfaces which are last
to collapse. As mentioned above, we will argue in section III that
it is possible that the size and separation of different collapsing  
domain walls may be of the order of a fraction of the horizon size just 
above the quark-hadron transition stage, i.e. of order of a km. 
If such large domain walls could form then the number of
baryons trapped inside can be very large. Also, due to larger mass
of the strange quark, reflection coefficient for them is larger than that
for the u and d quarks. This leads naturally to strangeness rich quark
nugget formation which, as we will show, can have baryon number as 
large as about 10$^{44}$ within a size of order 1 meter.

 In a previous paper we have shown that at the intersection of the three
different $Z(3)$ interfaces $l(x)$ vanish due to topological considerations,
leading to a topological string whose core is in the confining phase
\cite{znstr}. Structure of this string is similar to the standard axionic 
string which forms at the junction of axionic domain walls \cite{axion}. 
With quarks contributing to explicit $Z(3)$ symmetry breaking, this will
lead to decay of $Z(3)$ interfaces along with decay of the
associated strings. As $l(x) = 0$ in the core of these strings, collapsing
string loops will have larger reflection coefficients for quarks and will
also contribute to formation of baryon inhomogeneities. However, unless
this string network is very dense, large scale baryon inhomogeneities
will mostly result from collapsing $Z(3)$ interfaces.
                         
 The mechanism discussed in this paper will also lead to generation 
of baryon fluctuations in the QGP formed in relativistic heavy-ion 
collision experiments, with the walls forming during the initial
thermalization stage. The effects of explicit symmetry breaking
due to quarks on the evolution of wall etc., as mentioned above, 
will not be much relevant there because of very short time scale 
available for the evolution of QGP. We plan to study this using 
detailed computer simulations in a future work.

The paper is organized in the following manner. In section II we discuss 
structure of $Z(N)$ walls and give numerical results for the profile of 
$Z(3)$ walls for the case of QCD. Section III discusses how $Z(3)$ walls
can form in the early universe. In section IV baryon inhomogeneity
generation due to quark reflection from collapsing $Z(3)$ walls
is estimated. Numerical results and discussion are given in section V.

\section{structure of $Z(N)$ walls}

 We now start discussing the structure of $Z(N)$ interfaces.
We will first focus on pure SU(N) gauge theory and later
discuss the case with quarks. In this case, an order parameter for
the confinement-deconfinement phase transition is the Polyakov loop
$l(x)$ which is defined as,

\begin{equation}
l(x) = \frac{1}{N} tr \Bigl(P exp\Bigl( ig \int^\beta_0 A_0(x,\tau) 
d\tau \Bigr) \Bigr) .
\end{equation}

 Here $P$ denotes path ordering, $g$ is the gauge coupling, $\beta
= 1/T$, with $T$ being the temperature, $A_0(x,\tau)$ is the time
component of the vector potential at spatial position $x$ and Euclidean
time $\tau$. $l(x)$ is thus a complex scalar field. Under a global
$Z(N)$ symmetry transformation, $l(x)$ transforms as,
                                  
\begin{equation}
l(x) \rightarrow exp\bigl(\frac{2\pi i n}{N}\bigr) l(x),
~~ n = 0,1,..(N-1) .
\end{equation}  
 
 For temperatures above the critical temperature $T_c$, in the deconfining
phase, the expectation value of the Polyakov loop $l_0 = <l(x)>$ is non-zero
corresponding to the finite free energy of isolated test quarks. This
breaks the $Z(N)$ symmetry spontaneously. At temperatures below $T_c$,
in the confining phase, $l_0$ vanishes, thereby restoring the $Z(N)$
symmetry \cite{plkv}.

For making estimates, we will use the effective potential proposed by
Pisarski \cite{psrsk,psrsk2} (see, also ref. \cite{veff}) for the
Polyakov loop $l(x)$ for the case of QCD with $N = 3$. The effective
Lagrangian density is given by,

\begin{equation}
L = \frac{N}{g^2} |\partial_\mu l|^2 T^2 - V(l) .
\end{equation}

 Here, $N = 3$ and $V(l)$ is the effective potential for the Polyakov
loop given by,

\begin{equation}
V(l) = \Bigl( -\frac{b_2}{2}|l|^2 - \frac{b_3}{6}\bigl(l^3 + (l^*)^3
\bigr) + \frac{1}{4} \bigl(|l|^2 \bigr)^2 \Bigr) b_4 T^4 .
\end{equation}

 $l_0$ is then given by the 
absolute minimum of $V(l)$.  Values of various parameters in Eqs.(3),(4) 
are fixed in ref.\cite{psrsk2,dumitru1} by making correspondence to 
lattice results \cite{lattice}. Following \cite{dumitru1}, for three light 
quark flavors we take, $b_3 = 2.0$ and $b_4 = 0.6061 \times 47.5/16$, where 
the factor 47.5/16   accounts for the extra degrees of freedom relative 
to the degrees of freedom of pure gauge theory. $b_2$ is
taken as, $b_2(x) = (1-1.11/x)(1+0.265/x)^{2}(1+0.300/x)^{3}- 0.487$, 
where $x = {\rm T}/T_c$.  With the coefficients chosen as above, $l_0$ 
approaches the value $y =b_3/2 + \frac{1}{2}\sqrt 
{b_3^2 +4~b_2({\rm T}=\infty)}$ for temperature T $\rightarrow \infty.$ 
As in ref.\cite{psrsk2}, the fields and the coefficients are rescaled as 
$l \rightarrow l/y, b_2({\rm T}) \rightarrow b_2({\rm T})/y^2,b_3
\rightarrow b_3/y$ and $b_4 \rightarrow b_4~y^4$ to ensure proper
normalization such that the expectation value of the order parameter 
$l_0$ goes to unity for temperature T $\rightarrow \infty$.

By writing $l = |l| e^{i\theta}$ we see that the $b_3$ term in Eq.(4)
gives a $cos(3\theta)$ term, leading to $Z(3)$ degenerate vacua for
non-zero values of $l$, that is for T $> T_c$. The value of $T_c$ is 
taken to be $\sim 182$ MeV \cite{dumitru1}.  The $Z(3)$
interface solution will correspond to a planar solution (say in
the x-y plane) where $l$ starts at one of the minimum of $V(l)$ at
$z = - \infty$ and ends up at another minimum of $V(l)$ at $z = + \infty$.

  In our earlier work we have given profile of this
$Z(3)$ domain wall obtained by numerically minimizing the energy of
a suitably chosen initial configuration, see ref.\cite{znstr} for
details. Fig.1 gives the plot of $|l(z)|$ across the domain wall showing
the profile for the domain wall solution for T = 200 and 300 MeV. Note that 
the value of $|l(z)|$ in the middle of the wall is smaller for T = 200 MeV 
than for T = 300 MeV. We thus expect that the effective quark mass
will be larger for T = 200 MeV than for the case with T = 300 MeV inside the
wall leading to larger reflection coefficient for T = 200 MeV. The surface 
tension of the wall for T = 200 and 300 MeV are found to be about 0.34 
and 2.61 GeV/fm$^2$ respectively.  In an earlier 
work \cite{znstr} the surface tension was found to be about 7 GeV/fm$^2$
for T = 400 MeV. The values for T = 300 and 400 MeV are in reasonable 
agreement with the analytical estimates (for large temperatures) 
\cite{zntensn}.

\begin{figure}
\epsfig{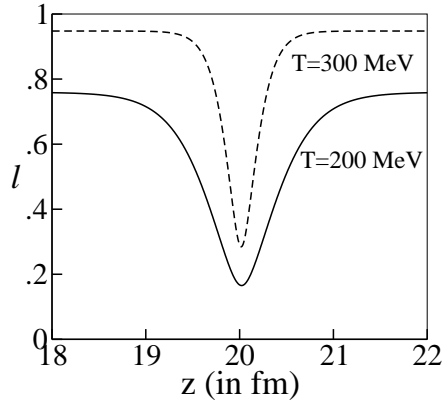} 
\caption{Profiles of the 
domain wall for T = 200 MeV (solid curve) 
and T = 300 MeV (dashed curve).} 
\label{fig:fig1}
\end{figure}

 Let us now come back to the issue of quarks and the $Z(3)$ symmetry.
The effect of quarks on this $Z(3)$ symmetry and $Z(3)$ interfaces etc.
has been discussed in detail in the literature \cite{qurk1,qurk2}. It has
been suggested that in the presence of quarks, the $Z(3)$ symmetry becomes
meaningless, and there is no sense in talking about $Z(3)$ interfaces etc.
\cite{qurk1}. It has also been advocated in many papers, that one can take
the effect of  quarks in terms of explicit breaking of $Z(3)$ symmetry
\cite{qurk2,psrsk,psrsk2}. In such a case, the interfaces will survive,
though they do not remain solutions of time independent equations of motion.
It has been argued in ref.\cite{psrsk2} that the effects of quarks in terms
of explicit symmetry breaking may be small, and the pure glue Polyakov model
may be a good approximation.  We will follow this interpretation, and
assume that the effect of quarks is just to contribute 
explicit symmetry breaking terms which can make the
interface and the string solution time dependent, but not invalid.
With the explicit symmetry breaking, the interfaces will start moving
away from the direction where true vacuum exists as in the conventional 
case of quark-hadron transition, and as mentioned above, for $Z(3)$ walls 
formed at some very early time in the universe (say near GUT scale), 
presumably all walls will disappear. This brings us to the issue of
the formation and evolution of these $Z(3)$ walls and strings which
we discuss in the next section. 

\section{formation of $Z(3)$ walls in the early universe}

The production of these $Z(3)$ walls and associated strings is, however, 
very different from the formation of conventional topological defects as
here the symmetry is broken in the high temperature phase, it gets
restored below $T_c$, the QCD transition temperature. To discuss the
formation of these objects, one can use the standard Kibble
mechanism \cite{kbl} invoking causality argument at a very early stage
of the universe. However, a concrete realization of the formation
of $Z(3)$ walls can be achieved in the context of inflationary models,
as we will discuss in this section.

 We mention that a scenario for the formation of $Z(3)$ domain
structure in the early universe has been discussed in ref.\cite{prsr1}
where it was proposed that a novel phase transition may occur in the 
universe at a temperature of order 10 TeV. The basic idea in ref.\cite{prsr1} 
is that if a large enough region was in a metastable $Z(3)$ vacuum of QCD  
initially then inflation can expand that region exponentially to
superhorizon size. The tunneling rate for decay of this superhorizon
size domain to the stable vacuum was estimated in ref.\cite{prsr1} 
(see also, ref.\cite{prsr2}) and it was concluded that bubble nucleation 
becomes effective when the universe temperature is around 10-20 TeV,
thereby leading to a new phase transition scale in the early universe.
However, as will be clear from the discussion below, a crucial ingredient 
in the model of ref.\cite{prsr1}, namely the assumption that such 
metastable domains survive the period of inflation, does not seem justified.

During inflation, the temperature of the universe is driven to almost
zero value due to rapid expansion. This will 
lead to barriers between different $Z(3)$ phases 
disappear when energy density drops below the QCD scale due to expansion, 
either in equilibrium, or out of equilibrium. One expects then that 
$l(x)$ will roll down to the unique minimum of the effective potential if 
the inflation time scale is larger than the roll down time scale (which 
should not be much larger than 1 fm at the stage, when the energy density
is of order of QCD scale, even for equations of motion in
the expanding background). This will happen if the inflation energy scale 
is below about 10$^9$ GeV, as in the low scale inflation models 
discussed later in this paper. This will lead to restoration of $Z(3)$ 
symmetry during inflation. $Z(3)$ symmetry will be subsequently broken
spontaneously as the universe reheats at the end of inflation to a 
temperature above $T_c$. $Z(3)$ domains and associated walls will then 
arise during this spontaneous symmetry breaking transition via the 
standard Kibble mechanism with typical sizes of the order of the 
correlation length at an appropriate stage during reheating (and 
therefore cannot have superhorizon sizes).
                                                   
 If the inflation time scale is much shorter than about  (10$^9$ GeV)$^{-1}$, 
then the condensate will not have time to roll down to the minimum of
the potential. It may be frozen (or might even decay during inflation), 
until reheating begins. (The nature of $l(x)$ in such non-equilibrium 
situations is not clear.  One can think of certain gauge field configurations 
which in equilibrium lead to appropriate behavior of $l(x)$, but simply get 
redshifted during inflation.) It seems natural to assume that the
potential energy of the condensate will be greatly reduced during inflation
as the relevant spatial region becomes devoid of matter by rapid expansion.
(With matter completely diluted away, the only relevant scales for this
potential energy can be the QCD scale, or quark masses). When reheating 
begins, universe gets filled with high energy particles from the decay of 
inflaton. The net energy density of this matter may then be very low just at 
the beginning of reheating, but it would not mean that the universe is getting 
heated from almost zero temperature upwards. Initially when the number of 
particles (from the decay of inflaton) is small, then mean free path of the 
particles will be larger than the Hubble size and the system will be 
completely out of equilibrium. As the density of these particles increases 
(and their energy decreases by multiple rescatterings), at some stage the
mean free path will become shorter than the Hubble scale and system can be
said to achieve (approximate) equilibrium. It seems clear that the energy
density content of this matter will be far greater than the potential energy
corresponding to any Polyakov loop condensate which could survive during
the inflationary stage. Therefore, the nature of resulting thermal 
quark-gluon system will be completely dominated by the decay products of 
inflaton, with any background Polyakov loop condensate possibly surviving 
through the inflation making negligible contribution to it. In other words,
the value of $l(x)$ (in equilibrium, or 
out of equilibrium) during reheating stage, and consequently, 
any resulting $Z(3)$ walls,  should be entirely determined by the newly 
created matter  and any memory of pre-existing Polyakov loop condensate 
will be lost. Therefore, in this case as well, one expects $Z(3)$ domain 
wall formation according to the Kibble mechanism, with typical sizes
of the order of relevant correlation length at an appropriate stage 
during reheating. (Same conclusions will be reached in models of 
preheating with parametric resonance.)
   
There are models of inflation where preheating can lead 
to a short secondary stage of inflation \cite{dblinfl} (see, also,
\cite{thrml}). However, the secondary inflation has short 
duration in these models, which seems inadequate to inflate
the $Z(3)$ domains to superhorizon sizes. Formation of truly 
superhorizon $Z(3)$ domains (as envisaged in ref.\cite{prsr1}) 
could be possible in the context of the so called {\it warm inflation} 
models where temperature does not become 
very low during inflation \cite{warm}. However the process of
inflaton decay during inflation is very complex in
these models \cite{warm2}. For example, at any stage, the thermal system 
consists of particles which have been freshly generated, as any previously
existing particles are diluted away by inflation. It is therefore
not clear if one can think of this as a pre-existing $Z(3)$ domain in 
equilibrium, with temperature changing during inflation. Instead, the
situation here appears to be closer to the case of high energy scale 
inflation, discussed above, where the matter is first diluted away, and 
then the space gets filled with completely new component of matter during 
reheating. Thus, even in warm inflation case, one may 
expect the behavior of $l(x)$, and hence $Z(3)$ domains, to be entirely
determined by the matter-radiation which is created near the end
of the inflation, leading to subhorizon $Z(3)$ domains.

  We therefore conclude that with generic inflationary models, one expects
formation of $Z(3)$ domains and associated walls (along with the 
strings \cite{znstr}) to arise during the $Z(3)$ 
symmetry breaking transition at the reheating (or preheating) stage after 
the inflation via the standard Kibble mechanism. For the evolution of this
domain wall (and associated string \cite{znstr}) network we note that the 
tension of the $Z(3)$  interface and this string is set by the QCD 
parameters and the temperature, hence their dynamics, as far as the tension 
forces are concerned, should be dominated by the background plasma (at
least by its QGP component) for temperatures far above the QCD scale.  
However, the explicit symmetry breaking due to quarks leads to pressure
difference between the metastable $Z(3)$ vacua and the true vacuum, and this 
should remain significant at high temperatures, again, because at high 
temperatures the only relevant scale is the temperature. As we mentioned above,
estimate of this pressure difference for high temperatures are given in 
ref.\cite{prsr1,prsr2}. (There have also been discussions of CP violating 
effects associated with the metastable phases \cite{cp}, such effects may be 
interesting in the context of our model). As mentioned above, due
to this pressure difference one expects that regions of metastable phases will
shrink quickly as walls enclosing the true vacuum expand. In this picture 
$Z(3)$ walls are unlikely to survive until late times, say until QCD scale,  
to play any significant role in the context of the universe. 

Though one may still not completely rule out the possibility
that the effects of explicit symmetry breaking due to quarks may not be
dominant at high temperatures so that walls may survive until late times. 
In this context we note that the wall motion at high temperatures should be
highly dissipative as quarks scattering from the walls will lead to
friction.  This is expected as the quark free energy depends on $l(x)$, 
hence there should be significant change in quark energy in crossing wall 
even at high temperatures (in a similar manner as discussed below), again, 
as T is the only relevant scale. For large friction the motion of wall in a 
local plasma rest frame will be strongly suppressed, with walls remaining
almost frozen in the plasma. For example, it has been discussed in the
literature that dynamics of light cosmic strings can be dominated by
friction which strongly affects the coarsening of string network
\cite{friction}.

However, we will discuss below a scenario where in the context of low 
energy scale inflationary models it is possible that large $Z(3)$ walls, 
with sizes of order of a fraction of the horizon size at the QCD scale 
may arise. In such a scenario,  with few large domain walls per 
horizon, the resulting inhomogeneities will be separated by large distances 
at the QCD transition scale (below which domain walls disappear as $l(x)$ 
becomes zero). With such large domain walls, number of baryons trapped 
inside can be very large.  As the reflection coefficient for the s quark 
is larger than that for the u and d quarks, it may also lead to strangeness 
rich quark nugget formation \cite{ngt} which can have baryon number as 
large as about 10$^{44}$ when walls collapse down to the size of order 1 
meter. Even if walls are not of such large sizes, still resulting baryon 
inhomogeneities may have large enough magnitudes and distance scales to be 
able to survive until nucleosynthesis and affect abundances of elements.
The model discussed in this paper can therefore be used to constrain
various models of low scale inflation using calculations of
inhomogeneous nucleosynthesis.

Recently inflationary models with low energy scale, near the
electroweak scale, have been proposed which satisfy various 
requirements for inflation \cite{ew}. These models have very low 
reheating temperature $T_{RH}$, which is below the electroweak scale, and 
can be as low as 1 GeV. Let us consider, in some detail,
formation of $Z(3)$ walls in the context of these models. At the end of
inflation the universe is almost at zero temperature before reheating
begins by the decay of the inflaton. As we discussed above, for inflation 
scales below about 10$^9$ GeV, this will lead to restoration of $Z(3)$
symmetry during inflation as $l(x)$ will have sufficient time to roll down 
to the unique minimum of the symmetry restored effective potential. 
Due to small coupling, decay of inflaton to other 
particles is very slow. However, due to very slow expansion rate of the 
universe near the electroweak scale, reheating still happens within one 
Hubble time in these models. (As opposed to high energy scale inflation
where the universe undergoes significant expansion during reheating. Also,
to keep our discussion simple, we are not discussing here the possibility
of preheating due to parametric resonance.)  We therefore have the situation
where the universe is slowly (compared to the universe expansion scale) 
heated from a low temperature up to the reheat temperature $T_{RH}$. As the 
temperature becomes larger than the quark-hadron transition temperature $T_c$, 
$Z(3)$ symmetry will be spontaneously broken and $Z(3)$ domain walls will 
appear.  (Note that the explicit symmetry breaking term can bias the formation
of $Z(3)$ domains as the temperature rises above $T_c$. We will assume
that thermal fluctuations, especially in view of continued heating by 
decay of inflaton, will dominate over any such bias.) Sizes of the resulting
$Z(3)$  domains, and hence of $Z(3)$ walls initially should depend on the 
details of reheating mechanism. For conservative estimates one may assume 
that these domains may not be much bigger than the QCD scale at the 
formation stage. (Though reheating, starting from a low temperature, 
may allow much larger coherence lengths leading to larger domains initially.)

For low scale inflationary model we are considering, evolution of the dense 
network of $Z(3)$ walls depends crucially on relative importance of tension 
and pressure forces. The estimates of ref.\cite{prsr1,prsr2} for pressure 
difference between the metastable $Z(3)$ vacua and the true vacuum are 
valid for high temperatures and hence are inapplicable here. This is why, 
even the decay rate for the metastable vacua as calculated in 
ref.\cite{prsr1,prsr2} cannot be used here. We can use the effective 
potential in Eq.(4), though it does not have explicit $Z(3)$ symmetry 
breaking term. Still, one can check from Eq.(4) that at, and near, $T_c$, 
the barrier between different $Z(3)$ vacua are much larger (by about a 
factor of 100) than the barrier between the broken and unbroken phase
\cite{znstr}, and the surface tension of $Z(3)$ walls remains
significant for temperatures near $T_c$. 

On the other hand it seems 
reasonable to assume that the pressure difference between the metastable 
$Z(3)$ vacua and the true vacuum resulting from the explicit symmetry breaking 
term may become very small near $T_c$ (see also, ref.\cite{psrsk2}). 
We will assume that this is the case. In such a case, the dynamics of $Z(3)$
walls near $T_c$  will be controlled by the surface tension of the walls,
with pressure difference remaining subdominant. This will also suppress decay 
of metastable phases by nucleation of true vacuum bubbles.  The evolution 
of a network of such walls will then be like the standard domain walls which
coarsens quickly and leads to few domain walls within the horizon volume.
For example, if we take the reheat temperature to be 1 GeV, then one
should get several large domain walls within the horizon while temperature
approaches the quark hadron transition temperature $T_c$. Important point here
is that during reheating stage, the temperature should remain near $T_c$ 
for large enough time so that the wall network can coarsen significantly
with pressure difference remaining subdominant. At the end of the reheating 
stage, with temperature reaching few GeV, pressure term should become 
important and walls should evolve depending on expansion rate and
wall velocity through the dissipative plasma. As mentioned above,
in view of large friction due to quark scatterings, wall velocity
may be very small and may help in retaining large sizes upto the stage
of quark-hadron transition.

 This scenario can lead to large $Z(3)$ domain walls at temperature near
the QCD scale. If pressure term starts dominating early, then domain
wall network may not be able to coarsen much and resulting walls will be 
smaller. Still resulting baryon inhomogeneities may have large enough 
scales to survive until nucleosynthesis and affect abundances of elements.  
In the optimistic scenario when temperature lasts
near $T_c$ for large enough time (depending on the details of
reheating mechanism) so that pressure remains subdominant, one may get 
almost horizon size walls at the final reheat temperature of
few GeV. Subsequent (dissipative) evolution of these walls, with expansion
of the universe stretching such large walls, one can get walls which 
have sizes of order of a fraction of the horizon size at QCD scale.
Also, as we mentioned above, there are models of inflation 
\cite{dblinfl,thrml,warm} in which larger domain walls can arise.

  We will assume such an optimistic scenario, and work out the consequences
of large $Z(3)$ domain walls near the QCD scale. As the walls evolve,
there will be volume contribution of energy coming from the explicit symmetry
breaking term. However, in the following calculations, we will neglect
these effects. This is because, as explained below, such effects will
require calculating reflection of quarks from a potential barrier which
depends on time (with temperature changing during wall collapse), which
will require much more elaborate simulations. 

\section{reflection of quarks from $Z(3)$ walls and baryon inhomogeneity
generation}
  
 To model the dependence of effective quark mass on $l(x)$ we could
use the color dielectric model of ref.\cite{phtk} identifying
$l(x)$ with the color dielectric field $\chi$ in ref.\cite{phtk}.
Effective mass of the quark was modeled in \cite{phtk} to
be inversely proportional to $\chi$. This leads to divergent quark mass
in the confining phase consistent with the notion of confinement.
However, we know that the divergence of quark energy in the confining
phase should be a volume divergence (effectively the length of string
connecting the quark to the boundary of the volume). $1/l(x)$ dependence
will not have this feature, hence we do not follow this choice. For the
sake of simplicity, and for order of magnitude estimates at this
stage, we will model the quark mass dependence on $l(x)$ in the following
manner.

\begin{equation}
m(x) = m_q + m_0(l_0 - |l(x)|)
\end{equation}

 This is somewhat in the spirit of the expectation that a linear term
in $l$ should arise from explicit symmetry breaking due to quarks
\cite{qurk2,psrsk,psrsk2}, though, as mentioned above, we are neglecting
the effects of explicit symmetry breaking between different $Z(3)$ vacua. 
Hence we use $|l(x)|$ in Eq.(5). Here $l(x)$ represents the profile of 
the $Z(3)$ domain wall, and $l_0$ is the vacuum value of $|l(x)|$ 
appropriate for the temperature under consideration. $m_q$ is 
the current quark mass of the quark as appropriate for the QGP 
phase with $|l(x)| = l_0$, with $m_u \simeq m_d = 10$ MeV and $m_s 
\simeq 140$ MeV. $m_0$ characterizes the constituent
mass contribution for the quark. We will take $m_0 = 300$ MeV. Note
that here $m(x)$ remains finite even in the confining phase with
$l(x) = 0$. As mentioned above, this is reasonable since we are
dealing with a situation where $l(x)$ differs from $l_0$ only in
a region of thickness of order 1 fm (thickness of domain wall). For
making conservative estimates, we will also give results for the
choice $m_0 = m_q$. This will lead to small value for the potential
barrier leading to small reflection coefficients. We will discuss
resulting baryon inhomogeneities for all these cases.  For very high
temperatures (e.g. for calculating friction for wall motion), one should 
use appropriate thermal masses. 

 Another simplifying assumption we make is to model the potential
barrier resulting from Eq.(5) as a rectangular barrier.
Height of the barrier $V_0$ is taken to be equal to $m(x) - m_q$ 
given in Eq.(5) with the smallest value of $l(x)$ in the profile of the 
domain wall (Fig.1).  The width of the barrier $d$ is taken to be equal to 
the width of the domain wall.  Using Fig.1, we take $d$ = 0.5 fm and 1 fm
for T = 300 MeV and 200 MeV respectively. Transmission coefficient $T$ 
for a quark of mass $m_q$ with energy $E$ for this  potential barrier can 
be straightforwardly calculated from the Dirac equation. We find:

\begin{equation}
T = {4r^2 \over 4r^2 + (1-r^2)^2 sin^2(p_2 d)}
\end{equation}

\noindent where,  $r = {p_2 (E + m_q) \over p_1 (E - V_0 + m_q)}$, $p_1^2 
= E^2 - m_q^2, {\rm and} ~~~ p_2^2 = (E-V_0)^2 - m_q^2$. For $|E-V_0| < m_q$, 
$p_2$  is imaginary and $sin^2(p_2 d)$ is replaced by $sinh^2(p_2 d)$.

 We now discuss the generation of baryon inhomogeneity.
We will assume that there are on the average $N_d$
domain walls per horizon volume and will present results for
$N_d = 1$ and $N_d = 10$. As the walls collapse, there 
will be some reheating from decreasing surface area, and from
explicit symmetry breaking due to quarks. However, we will neglect these 
effects in the present discussion, so that we can use a fixed potential 
barrier (corresponding to a fixed temperature) for calculating baryon 
transport across the wall.   We will also assume that wall collapse 
is rapid, say with a velocity $v_w$ equal to the velocity of 
sound $c/\sqrt{3}$ (it could be larger
if wall tension completely dominates over the friction). In this case,
walls should collapse away in a time smaller than the Hubble time. Thus 
for rough estimates, one can neglect the expansion of the universe while 
studying the collapse of a single domain wall (in contrast to 
ref.\cite{fuller,layek}). Again, this has the simplification that one can use
a fixed shape for the potential barrier, appropriate for a fixed temperature.
As a fraction of quarks and antiquarks is reflected by the collapsing
wall, thermal equilibrium should be maintained as in the conventional case 
\cite{fuller}. This will lead to concentration of net baryon density 
inside such that we can use the transmission coefficient (Eq.(6)) for the 
net baryon number.

 We mention here that in the context of heavy-ion
collisions this assumption of rapid equilibration of reflected quarks and
antiquarks may not hold true. In that case, the concentration of strange
quarks as well as antiquarks may build up inside the collapsing walls which
can lead to important effects such as enhancement of strange hadrons etc. 

 Let us denote by $n_i$ and  $n_o$ the net baryon densities 
in quarks in the region inside the collapsing domain wall (with volume $V_i$), 
and the region outside of it (with volume $V_o = V_T - V_i$) respectively. 
$V_T$ is the total, fixed, volume of the region neglecting the 
expansion of the universe as discussed above. 
Total baryon numbers are then given by $N_i = n_i V_i$ 
and $N_o = n_o V_o$ for inside and outside regions respectively.
The evolution equations for $n_i$ and $n_o$ can be 
written as follows (by straightforward modification of the approach
used in \cite{fuller,layek}),

\begin{eqnarray}
\dot n_i =  [- {2 \over 3} v_w T(v_w) n_i + {n_o T(v_q^-) -
n_i T(v_q^+) \over 6}] {S \over V_i} - n_i {{\dot V_i} \over V_i} \\
\dot n_o =  [ {2 \over 3} v_w T(v_w) n_i - {n_o T(v_q^-) - n_i 
T(v_q^+) \over 6}] {S \over V_o} + n_o {{\dot V_i} \over V_o}
\end{eqnarray}

\noindent Here dot denotes the time derivative and $S$ is the surface 
area of the domain wall.  $T(v_w)$ is the transmission 
coefficient for quarks which have thermal velocity 
parallel to the domain wall (with corresponding number density being
$4n_i/6$). This is calculated by using Eq.(6) for the relative 
velocity $v_w$ between the quark and the wall.  $T(v_q^+)$  and
$T(v_q^-)$  are transmission coefficients for quarks with thermal 
velocities towards the wall from inside  and from outside respectively 
(corresponding densities being $n_i/6$ and $n_o/6$),
calculated with appropriately Lorentz boosted energies. 
At these temperatures, the thermal velocities $v_q$ of u,d,s quarks
will be close to the speed of light $c$. (We mention here that
the explicit symmetry breaking between different $Z(3)$ vacua will
also lead to asymmetry in the transmission coefficients from the
two sides of the wall. Though, for large enough potential barrier
this difference may not be very significant, especially near $T_c$.) 

 The volume enclosed by the spherical collapsing wall is $V_i(t) = 
{4\pi \over 3} R(t)^3$ with the radius $R(t)$ given by

\begin{equation}
R(t) = {r_H \over 2 N_d^{(1/3)}} - v_w (t - t_0)
\end{equation}
                                                                                
\noindent where $r_H (= 2t)$ is the size of the horizon at the 
initial time $t_0 \simeq 30 ({150 \over {\rm T(MeV)}})^2 ~~\mu$sec.  
We take fixed volume $V_T = r_H^3/N_d$ as appropriate for a single
collapsing domain wall.  With $R(t)$ given by Eq.(9), one has to solve 
Eqs.(7),(8) simultaneously to get the detailed evolution of baryon density in
the region enclosed by the collapsing domain wall.  Baryon inhomogeneity
will be produced as baryons are left behind the collapsing wall. We mention
here that during the final stages of collapse of domain wall, baryon
overdensities may be so large that the chemical potential becomes
comparable to the temperature. This will have to be taken into
account when calculating the reflection of quarks from the collapsing
walls. However, we do not study the evolution of overdensities during
those final stages, hence we can neglect the effects of the chemical
potential. 

For the profile of the baryon inhomogeneities, if $\rho (R)$ is the 
baryon density left behind at position $R$ from the center of the 
collapsing spherical wall, then $N_i (R+dR) - N_i (R) =  \rho (R) 
4\pi R^2 dR$.  With the time dependence of $R$  given above, we get,

\begin{equation}
\rho(R) = {dN_i \over dR} {1 \over 4\pi R^2} = 
- {{\dot N_i} \over 4\pi v_w R^2}  ~~~~.
\end{equation}

We mention here that the derivation of Eq.(10)  assumes that baryons left
behind by the collapsing interfaces do not diffuse away, while the 
derivation of equations for baryon transport across the wall (Eqs.(7),(8)) 
assumed that baryons in both regions homogenize, so that those equations
could be written only in terms of two baryon densities, one for each 
region \cite{fuller,layek}.  A more careful treatment should take 
proper account of baryon diffusion.

  Eqs.(7),(8) are numerically solved simultaneously to 
get the evolution of baryon densities $n_i$, and $n_o$.  We have normalized 
the initial densities to the average baryon density of the universe $n_{av}$
at that temperature. Initial values of $n_i$ and $n_o$ are thus equal to 1. 
We have checked that the total baryon number $N_i + N_o$ remains almost
constant in time. We find that there are very small random fluctuations 
in the value of total baryon number, with no tendency of net increase or
decrease over time. Numerical errors are therefore under control.
Resulting profiles of baryon overdensity $\rho(R)$ is calculated
using Eq.(9) and Eq.(10).  We have used Mathematica routines for
numerically solving these coupled differential equations.

 Evolution of baryon inhomogeneities of varying amplitudes and  length
scales has been analyzed in detail in literature \cite{bfluct2}. From ref.
\cite{bfluct2} one can see that baryon inhomogeneities of initial
magnitude $n_i/n_{av} \sim 1000$ near the QCD scale should survive
relatively without any dissipation until the nucleosynthesis stage when 
temperature T $\sim 1$ MeV for all the values of length scales relevant for us,
i.e. few tens of cm and above. (For example inhomogeneities with
baryon to entropy ratio of about $10^{-5}$ almost do not change during
their evolution. Inhomogeneities with larger amplitude eventually
dissipate to this value. See, ref.\cite{bfluct2}.)  Though, the length
scales in ref.\cite{bfluct2} are taken to be comoving at 100 MeV, the
results there should apply for the order of magnitude estimates
for the values of temperature we have considered T $\simeq 200$ MeV.
Also, as these inhomogeneities in our model are produced above the
quark-hadron transition, they may affect the quark-hadron transition
dynamics \cite{soma}. As discussed in ref.\cite{soma}, modified
dynamics of transition can lead to amplification of these already
formed overdensities.

\section{Results and discussion}

 In Fig.2 we have given plots of $n_i$ vs. time (in microseconds)
and of $\rho$  vs. R (in meters) for T = 200 MeV and for the choice
of $m_0$ = 300 MeV in Eq.(5).  (Again, initial values
of $n_i$, $n_o$ are normalized to the average baryon density of the
universe $n_{av}$. To get absolute values of these densities, and of
$\rho$, one should multiply by $n_{av}$.) We have taken
the number of domain walls in a horizon volume $N_d$ to be 10.
We find that the size of the region inside
which the baryon overdensity $\rho > 1000$ is about 10 m for u,d quarks
while the size is about 60 m for the strange quark case.
Baryon density sharply rises for small $R$.  We see that for $R < 1 m$,
$\rho$ rises to a value of about 20,000 for u,d quarks and to a value of
about 6 $\times 10^5$ for the strange quark. These overdensity
magnitudes and sizes are large enough that they can survive until
the time of nucleosynthesis and affect nuclear abundances.
Typical separation between the inhomogeneities is the inter-domain wall
separation near the QCD scale (below which walls disappear), and hence 
can be very large in our model, of order of a km. (Of course, with the
assumption that large size walls arise at the end of reheating stage in
a low scale inflationary model, as discussed in section III.) 
This corresponds to about 100- 200 km 
length scale at the nucleosynthesis epoch, which is  precisely the range 
of length scales which can have optimum effects 
on nucleosynthesis calculations  in ref.  \cite{ibbn2}.

\begin{figure}
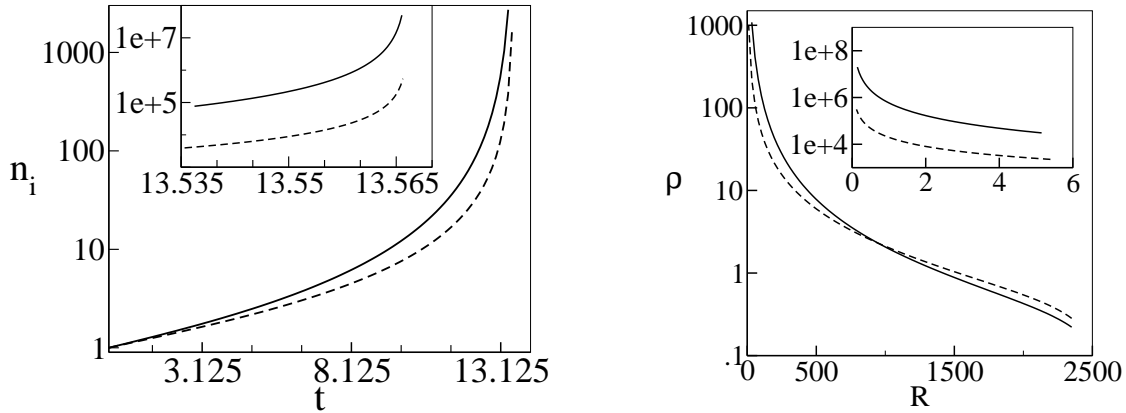

\epsfig{file=fig2a.eps,height=65mm} 
\epsfig{file=fig2b.eps,height=53mm} 
\caption{Plots of $n_i$ vs. time $t$ (in microseconds), and 
$\rho$  vs. R (in meters). The origin for $t$ is chosen at the 
beginning of the wall collapse.  Solid curves are for s quark and
dashed curves for u,d quarks.} 
\label{fig:fig2}
\end{figure}

 When we consider only one domain wall in the horizon ($N_d$ = 1) then
overdensities are larger. For example for the above cases, we find 
that within $R < 1$ m, $\rho$ is larger by a factor of 2 to 4.
Overdensities become much smaller for the u,d quark case if we
take $m_0$ in Eq.(5) to be equal to $m_q$ (instead of 300 MeV), as
the potential barrier becomes much smaller than the typical quark
energy leading to very small reflection coefficient. For example, 
for other parameters of Fig.2, $\rho$ is about 20
for $R < 1$ m for u,d quark. However, for the strange quark
even with $m_0 = m_q$ the potential barrier is high enough 
with significant reflection of quarks and leads to $\rho = 120000$
for $R < 1$ m. For comparison we have also calculated overdensities
occurring at T = 300 MeV. These are much smaller, first due to smaller
domain wall width, and secondly due to larger value of $l$
in the domain wall (see, Fig.1), leading to smaller potential
barrier (height as well as width). For example, with
$m_0$ = 300 MeV, within $R < 1$ m we get $\rho = 5000$ for $s$ quark, 
and $\rho = 400$ for u,d quarks.

  With large overdensities occurring as in Fig.2, there may be possibility
of quark nugget formation \cite{ngt}. Indeed we find that for certain cases,
e.g. with the parameters of Fig.2, total number of baryons can be very
large, $\sim 10^{44}$ within $R = 1$ m. These regions will be dominated by
strange quarks as is clear from Fig.2. These seem to be favorable
conditions for the formation of stable quark nuggets. If these 
survive cooling down through $T_c$, and survive until present then 
they may constitute dark matter, without affecting microwave background 
anisotropy or nucleosynthesis constraints. 
 
  We summarize main features of our model. We have discussed formation
and evolution of $Z(3)$ domain walls in the early universe. We have
argued that, in the context of low scale inflationary models with reheat
temperature of order of few GeV, it is
possible that large $Z(3)$ walls can arise near the QCD scale. (We also 
briefly mentioned other possibilities where large $Z(3)$ walls can arise in 
inflationary models based on thermal inflation, or warm inflation etc.)
We study baryon inhomogeneities resulting from these walls.  In our model, 
baryon inhomogeneities are produced not due to moving quark-hadron 
phase boundaries as in the conventional treatments, but due to moving 
$Z(3)$ domain walls. The variation in the value of the Polyakov loop 
order parameter across the wall leads to non-zero 
reflection coefficient for the quarks. As a closed domain 
wall collapses, a fraction of quarks inside it remains trapped
leading to production of baryon inhomogeneities.
Important thing is that all this happens in the QGP phase
itself, with any possible quark-hadron transition being completely
irrelevant.  We have assumed that near $T_c$, the pressure difference 
between the metastable $Z(3)$ vacua and the true vacuum may be small so that 
surface tension may play a dominant role in the early evolution of domain 
walls, which form as the temperature of the universe crosses $T_c$ during 
reheating stage at the end of inflation.  The separation of the resulting 
inhomogeneities is then the separation between different collapsing  domain 
walls, which may be of the order of a fraction of the horizon size
near the quark-hadron transition stage. Resulting 
overdensities then have large enough magnitudes and sizes that they can 
survive until the stage of nucleosynthesis and affect the abundances of 
elements. We also find that if such large walls can form then strangeness 
rich quark nuggets of large  baryon number ($~ 10^{44}$) can form in our 
model.  If the effects of pressure difference do not 
remain subdominant near $T_c$ in the coarsening dynamics due to 
surface tension of walls, then resulting walls will not be as large. 
Still resulting baryon inhomogeneities may have large enough scales 
to survive until nucleosynthesis and affect abundances of elements.  
In view of tight constraints on models of inhomogeneous nucleosynthesis,
our results can be used to constrain various models of low scale
inflation (or other inflationary models, as discussed above). 

 The mechanism discussed in this paper will also lead to generation 
of baryon fluctuations in the QGP formed in relativistic heavy-ion 
collision experiments, with the walls forming during the initial
thermalization stage. The effects of explicit symmetry breaking
due to quarks, as discussed above, will not be much relevant there 
because of very short time scale available for the evolution of QGP.
However,  one cannot use simplifying assumptions
about coarsening of $Z(3)$ walls for the heavy-ion case, as one can do 
for the case of the universe. Similarly, because of rapid cooling due to
expansion, one will have to use time dependent potential barrier for
estimating quark reflection from $Z(3)$ walls.  We plan to study this 
using detailed computer simulations in a future work.

\section*{Acknowledgments}

  We are very thankful to Krishna Rajagopal for very useful discussions and 
comments, especially for emphasizing to us the importance of explicit symmetry
breaking due to quarks at high temperatures. We are also grateful to Sanatan 
Digal, Balram Rai and Rajarshi Ray for very useful comments and suggestions 
about the paper.  AMS and VKT acknowledge the support of the Department
of Atomic Energy- Board of Research in Nuclear Sciences (DAE-BRNS),
India, under the research grant no 2003/37/15/BRNS.


\end{document}